%% file: IRAS18151_final.tex
\documentclass{emulateapj}

\input{acronyms.tex}

\usepackage{graphicx}
\usepackage{subfigure}  
\usepackage{natbib}
\usepackage{color} 
\usepackage{txfonts}

\shorttitle{IRAS 18151-1208}
\shortauthors{Fallscheer et al.}

\begin{document}

\title{A high mass dusty disk candidate: the case of IRAS 18151-1208}

   \author{C. Fallscheer\altaffilmark{1} and H. Beuther} 
        \affil{Max Planck Institute for Astronomy, K\"{o}nigstuhl 17, 69117 Heidelberg, Germany}
	\email{Cassandra.Fallscheer@nrc-cnrc.gc.ca}
   \author{J. Sauter\altaffilmark{2} and S. Wolf}
	\affil{University of Kiel, Institute of Theoretical Physics and Astrophysics, Leibnizstrasse 15, 24098 Kiel, Germany}
   \and
   \author{Q. Zhang}
	\affil{Harvard-Smithsonian Center for Astrophysics, 60 Garden St., Cambridge, MA 02138, USA}

\altaffiltext{1}{Current affilliation: National Research Council, Herzberg Institute of Astrophysics, 5071 West Saanich Rd., Victoria, BC, V9E2E7, Canada}
\altaffiltext{2}{Also at: Max Planck Institute for Astronomy, Heidelberg, Germany}

\begin{abstract}
Many questions remain regarding the properties of disks around massive prototstars.  
Here we present the observations of a high mass protostellar object including an elongated dust continuum structure perpendicular to the outflow. Submillimeter Array 230 GHz line and continuum observations of the high mass protostellar object IRAS 18151-1208 along with single dish IRAM 30\,m observations afford us high spatial resolution (0.8$\arcsec$) as well as recovery of the extended emission that gets filtered out by the interferometer. The observations of $^{12}$CO confirm the outflow direction to be in the southeast-northwest direction, and the 1.3 mm continuum exhibits an elongation in the direction perpendicular to the outflow.  We model the physical parameters of the elongated structure by simultaneously fitting the observed spectral energy distribution (SED) and the brightness profile along the major axis using the 3D Radiative Transfer code MC3D.  Assuming a density profile similar to that of a low mass disk, we can also reproduce the observations of this high mass protostellar object.  This is achieved by using the same density distribution and flaring parameters as were used in the low mass case, and scaling up the size parameters that successfully modeled the circumstellar disk of several T Tauri stars.  We also calculate that a region within the inner 30 AU of such a high mass disk is stable under the Toomre criterion. While we do not rule out other scenarios, we show here that the observations in the high mass regime are consistent with a scaled up version of a low mass disk. Implications on high mass star formation are discussed.
\end{abstract}


\section{Introduction}

In contrast to low mass T Tauri stars and even intermediate mass Herbig stars of which numerous examples exist where circumstellar disks have been detected (see review by \citealt{wats2007}), the role of disks in the case of massive star formation is not yet well defined.  In the low mass case, it is clear that accretion disks and molecular outflows are integral parts of the formation process. While the occurrence of massive molecular outflows is well established in cases of massive star formation, evidence for the analogous disk component has only tentatively been identified (see review by \citealt{cesa2007}). 

In a concerted effort to shed some light on accretion disks in high mass star formation, we observed a sample of seven candidates with the Submillimeter Array (Zhang et al. \emph{in prep}).  The high mass protostellar object IRAS 18151-1208 is one of the sources included in this study.  This source was also included in the single dish survey of IRAS sources conducted at the IRAM 30\,m telescope by \citet{srid2002,beut2002hmpo}, and has also been observed with the IRAM 30\,m and Mopra single dish telescopes by \citet{mars2008}.  IRAS 18151-1208 is at a distance of 3.0 kpc  and has a luminosity of approximately 10$^{4}$ L$_{\odot}$ \citep{srid2002}.  

Previous single dish observations \citep{beut2002outflow} suggested the presence and general morphology of outflow in the region.  However, data quality was not good enough to observe specific details.  Narrow-band H$_2$ ($\lambda$ = 2.122 $\mu$m) observations \citep{davi2004} of IRAS 18151-1208 show evidence of two collimated molecular outflows originating from two separate sources. 
In the case of low mass star formation, it is clear that accretion disks play a central role in the existence of outflows.  
However, in the high mass regime, observations of scaled up versions of these low mass accretion disks are rare \citep{cesa2007}.  
Our goal is to show that  
a scaled-up version of a disk modeling scheme that successfully reproduces the observations of disks around T Tauri stars 
can be applied in the case of a massive protostellar object, namely IRAS 18151-1208.

The T Tauri star IRAS 04302+2247, nicknamed the ``Butterfly Star'' because of its butterfly-shaped morphology, 
has a well-defined circumstellar disk.  \citet{wolf2003,wolf2008} model the circumstellar environment of this source using a Monte Carlo 3D (MC3D) Radiative Transfer code. 
Similarly, the disk in the Bok globule CB 26 \citep{saut2009} has also been observed and modeled.  CB 26 is also a T Tauri star 
destined to become a low mass star with an edge-on circumstellar disk.  Additionally, a handful of other edge-on disks around T Tauri stars have been modeled using a similar technique [e.g. HK Tau \citep{stap1998}, HV Tau \citep{stap2003}, and IM Lupi \citep{pint2008}].  On the theoretical side, \citet{whit2003} use a similar Monte Carlo radiative transfer approach to test various envelope/disk geometries for Class I sources.  While theoretical descriptions such as \citet{whit2003} and observational studies (\citealp{sici2006}, for example) of low mass T Tauri disks are relatively well developed, such a complete picture is missing in the high mass regime. 



\section{Observations}

 \begin{table*}
 \caption{SMA observation parameters.  Entries include the date of observation, the array configuration, the calibrators, the synthesized beam obtained after inverting and cleaning the data, and the 1$\sigma$ continuum noise level. 
\label{obs_params}}
 \centering 
 \begin{tabular}{c|c|c|c c|c|c}
 \tableline\tableline
 Date & Freq & Configuration 	& \multicolumn{2}{c|}{Calibrators} & beam & rms$_{cont}$ \\ 
      & [GHz]  & 			& Bandpass & Phase \& Amplitude	& [$''$] 	& [mJy/bm] \\ 
 \tableline
 \tableline  
 2007 May 17 & 230 & very extended & 3C273  & 1743-038 \& 1911-201  & & \\ 
 2007 Jul 08 & 230 & compact       & 3C273  & 1743-038              & & \\
\tableline
             & 230 & combined      &        &                       & 0.8x0.7 & 2 \\ 
 \tableline
 \end{tabular}
 \end{table*}

\begin{table}
\caption{Observed lines \label{obs_lines}}
\centering
\begin{tabular}{lcr}  
\tableline\tableline
Transition 				& Rest Frequency & E$_{upper}$ 	\\    
					& [GHz] 	& [K]	\\
\tableline
C$^{18}$O 2 $\rightarrow$ 1		& 219.560	& 16	\\
SO 6$_5$ $\rightarrow$ 5$_4$		& 219.949       & 35    \\
$^{13}$CO 2 $\rightarrow$ 1		& 220.399	& 16	\\
CH$_3$OH 8$_{-1}$ $\rightarrow$ 7$_0$ E	& 229.759	& 89	\\
CO 2 $\rightarrow$ 1			& 230.538	& 17	\\
\tableline
\end{tabular}
\end{table}

\begin{figure}
\includegraphics[angle=-90,scale=0.7]{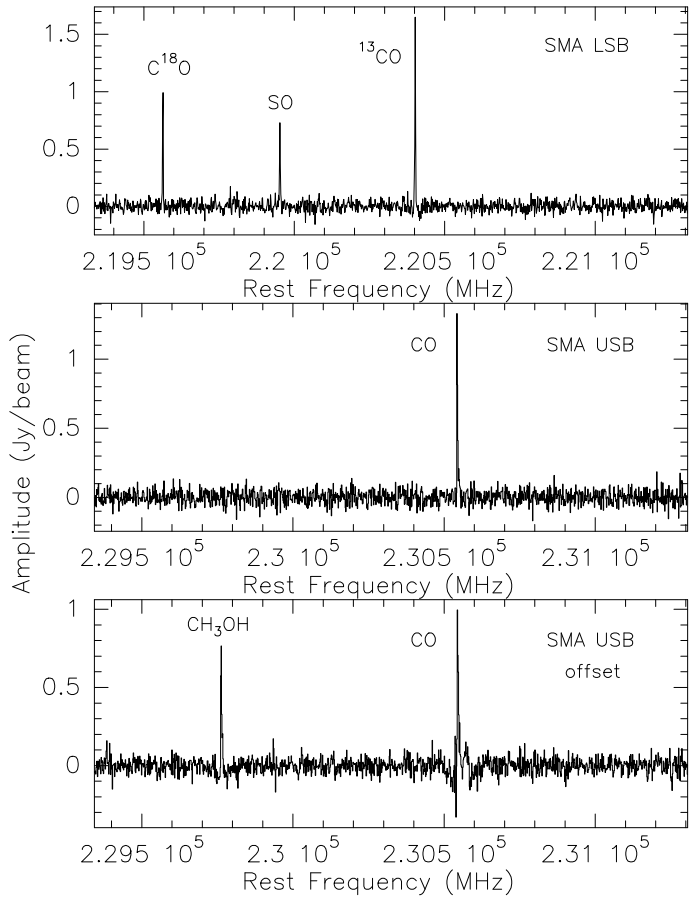}
\caption{The spectra from the SMA.  \emph{Upper:} the lower side band spectrum at the center of the 1.3 mm dust continuum peak of the combined compact and very extended configuration data. \emph{Middle:} the same as above except for the upper side band. \emph{Lower:} the upper side band at a position approximately 7$\arcsec$ northeast of the continuum peak only including the compact configuration data. \label{spectra}}
\end{figure}

\subsection{Submillimeter Array}
IRAS 18151-1208 was observed at 1.3 mm with the Submillimeter Array (SMA)$\footnote{The Submillimeter Array is a joint project between the Smithsonian Astrophysical Observatory and the Academia Sinica Institute of Astronomy and Astrophysics, and is funded by the Smithsonian Institution and the Academia Sinica.}$ in the compact and very extended configurations on 2007 July 8 and 2007 May 17 respectively.  Observations in the compact array were made under stable and excellent weather conditions, with a $\tau$(230 GHz) of approximately 0.07.  The very extended configuration observations were made under observing conditions with a $\tau$(230 GHz) of 0.12.  The combination of these two configurations provided baseline lengths varying between 14 and 520 meters corresponding to 10 k$\lambda$ and 400 k$\lambda$ at 1.3 mm.  The phase reference used was RA(J2000) = 18$^{h}$17$^{m}$57.1$^{s}$ and Dec(J2000) = -12$^{\circ}$07$\arcmin$22$\arcsec$ with the emission peak approximately 17$\arcsec$ east of these coordinates.  We adopted a rest velocity v$_\mathrm{LSR}$ = 32.8 
km s$^{-1}$ \citep{srid2002}.

For both data sets, the quasar 3C273 was used as the bandpass calibrator, and in between each 10 minute source observation, 5 minute observations of the quasar 1743-038 were made for phase and amplitude calibration.  The quasar 1911-201 was used as an additional phase and amplitude calibrator in the very extended data set.  Fluxes for the calibration sources were obtained from the Submillimeter Calibrator List$\footnote{ http://sma1.sma.hawaii.edu/callist/callist.html}$.  Measured fluxes after the calibration were consistent with the tabulated values to within 20\%.  
A summary of this information can be found in Table~\ref{obs_params}.

For the mapping, a miriad robust parameter of 0 was used for the continuum while a more natural weighting scheme (larger values of the robust parameter) was used for the line data. After combining the SMA data sets, the synthesized beam size of the continuum is 0.8$\arcsec\times$0.7$\arcsec$. A summary of this and other observation parameters are given in Table~\ref{obs_params}.  Spectral resolution of 0.55 km s$^{-1}$ was obtained over a 4 GHz bandwidth divided into a lower and upper sideband separated by 10 GHz (see Fig.~\ref{spectra}) with central frequencies of 219.6 and 230.5 GHz respectively. The lines observed by the SMA are listed in Table~\ref{obs_lines}.  These are discussed in more detail in Section \ref{linedata}.

\subsection{Pico Veleta 30\,m Telescope} \label{30m}
The shortest baseline in our SMA datasets is 14 m 
which corresponds to a spatial scale of 19$\arcsec$.  The SMA is not sensitive to structures larger than this, so supplementary short spacing information was obtained in the on-the-fly mode with the HERA receiver on the IRAM 30\,m telescope near Granada, Spain.  These observations were made on 2006 November 12 under very good observing conditions with $\tau$(230) of 0.07.  The receiver was tuned to 230.5 GHz centered on the $^{12}$CO(2--1) line.  A spectral resolution of 0.4 km s$^{-1}$ was obtained with a channel spacing of 0.3 MHz. 
The spectra were first processed with CLASS90 of the GILDAS package, then 
further analyzed with GREG.  While the single dish data can in principle be combined with the interferometer data, 
their quality is not good enough to do so in this case, so we analyze the data sets independently.  
\section{Observational Results}

\subsection{Millimeter Continuum Emission}\label{mmcontinuum}

\begin{figure*}
\centering
\includegraphics[angle=-90,scale=0.5]{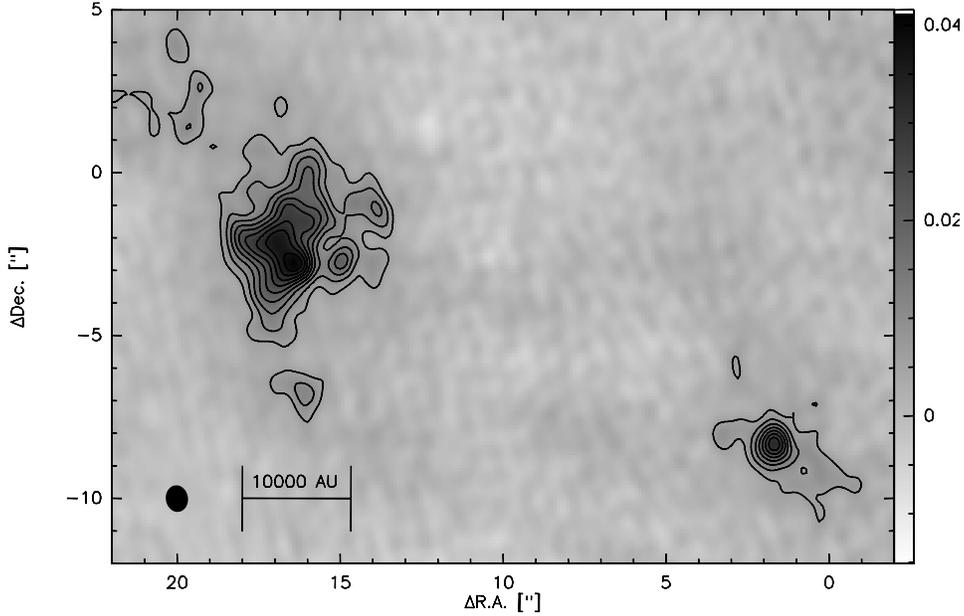}
\caption{The 1.3 mm dust continuum.  The size of the beam is indicated in the lower left corner.  Contours start at 3$\sigma$ increasing in steps of 2$\sigma$ where $\sigma$ is 2 mJy/beam. \label{dustcont}}
\end{figure*}

\begin{figure}
\includegraphics[angle=-90,scale=0.55]{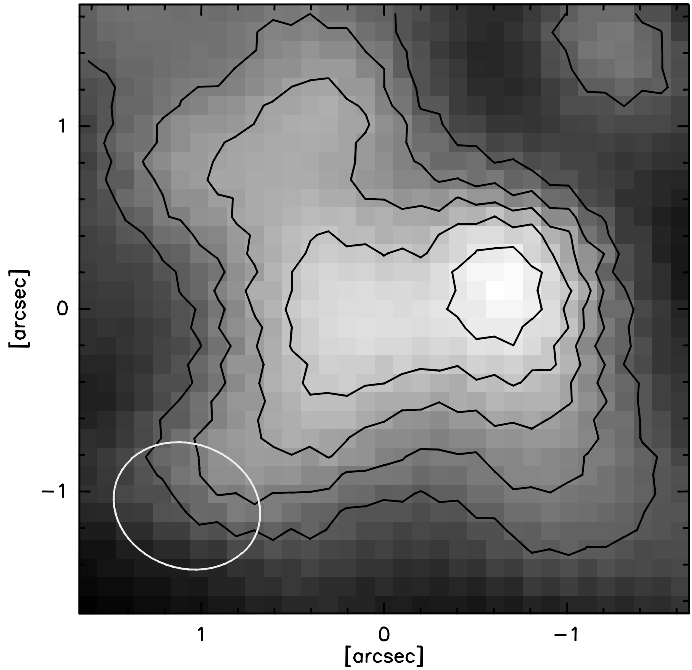}
\caption{A zoomed in view of the primary peak of the 1.3 mm dust continuum.  Here we have rotated the image by 61$^{\circ}$ in order to align the elongation with the horizontal axis.  The size of the beam is indicated in the lower left corner. The integrated flux in this area is 0.4 Jy.  One arcsecond corresponds to 3000 AU. \label{contzoom}}
\end{figure}

\begin{table} 
\caption{Position angles (measured East of North) of the components associated with the source. \label{pos_ang}}
\centering
\begin{tabular}{cc}
\tableline
\tableline
Component & PA \\
  &  $\circ$   \\
\tableline
H$_2$ jet & 128 \\
SMA $^{12}$CO outflow &  130\\
30\,m $^{12}$CO outflow & 108 \\
continuum elongation & 29 \\
 
\tableline
\end{tabular}
\end{table}

\begin{figure*}
\includegraphics[angle=-90,scale=0.9]{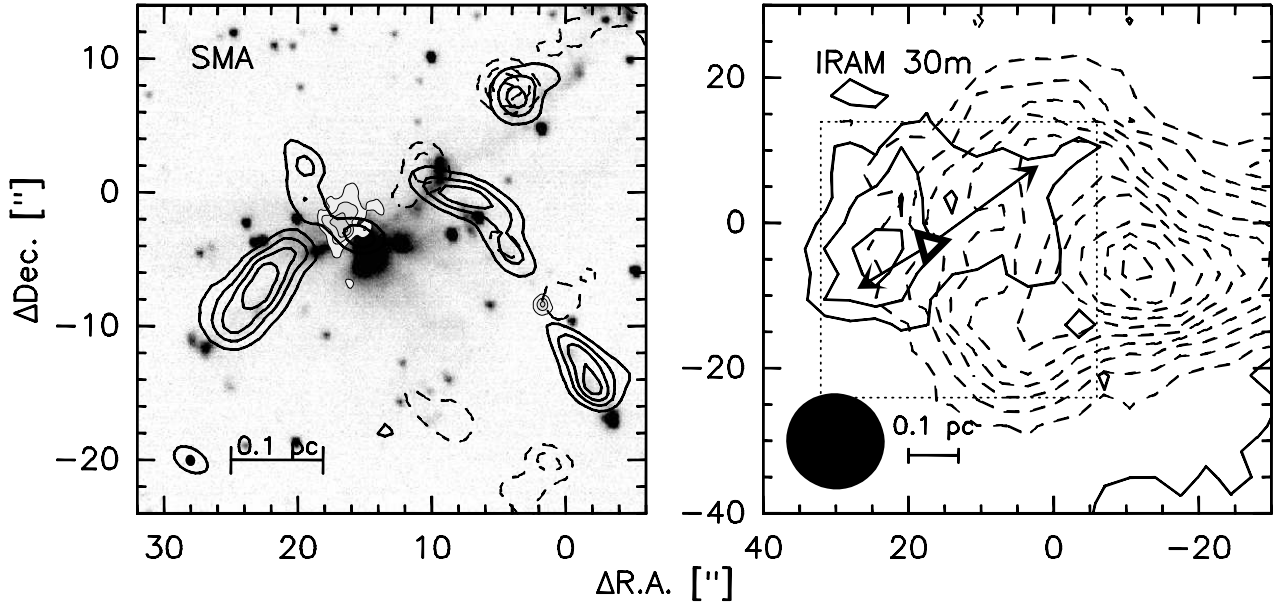}
\caption{The outflow in $^{12}$CO(2--1).  Integrated maps of IRAS 18151-1208 over the entire velocity range of the line. Blueshifted emission (thick solid contours) integrated over velocities of 23-29 km s$^{-1}$ and redshifted emission (thick dashed contours) integrated over velocities of 40-50 km s$^{-1}$. 
\emph{Left:} SMA $^{12}$CO(2--1) and continuum (grayscale with thin solid contours) overlaid on the H$_2$ map of \citet{davi2004}.  Contours for CO start at 3$\sigma$ and increase in steps of 3$\sigma$ where $\sigma$ is 0.15 and 0.16 Jy bm$^{-1}\cdot$kms$^{-1}$ for the blue- and redshifted emission respectively. \emph{Right:} IRAM 30\,m $^{12}$CO(2--1).  The 1.3 mm dust continuum peak is indicated by the triangle, and the outflow direction and extent as determined by the SMA data are indicated by arrows. The region plotted in the SMA figure at left is indicated by the dotted lines. Contours start at 3$\sigma$ and increase in steps of 3$\sigma$ where $\sigma$ is 11.4 Jy bm$^{-1}\cdot$kms$^{-1}$ for the blueshifted component and 10.4 Jy bm$^{-1}\cdot$kms$^{-1}$ for the redshifted component. \label{outflow}}
\end{figure*}

Combining the very extended and compact configuration interferometer data, we map the 1.3 mm dust continuum at a spatial resolution of $\sim$2500 AU (Fig.~\ref{dustcont}). Aside from the primary peak, we detect a secondary source 16$\arcsec$ to the southwest of the primary.  Zooming into the primary peak (Fig.~\ref{contzoom}) displays the elongation of the dust continuum in greater detail.  This elongation is perpendicular to the bipolar molecular outflow shown in Fig.~\ref{outflow}, and may be associated with a disk component in a roughly edge-on orientation.  The position angles of the relevant components are listed in Table \ref{pos_ang}. To simplify the analysis of the modeling, we have rotated the zoomed in image by 61$^{\circ}$ such that the elongation lies along the horizontal axis.  The brightness profile along this axis is asymmetric, with the difference in height of the two peaks on the order of 1$\sigma$ (see Fig.~\ref{rad_den}).

We measure an integrated flux of 0.4 Jy in the region contained within the 10,000 AU x 10,000 AU box shown in Fig.~\ref{contzoom} and a peak flux of 0.042 Jy.  Following the methods of \citet{hild1983} and \citet{beut2002outflow, beut2005} we adopt a dust opacity index $\beta_{\rm op}$ of 2.35 \citep{wein2001} which corresponds to the dust model used in our modeling described in Section \ref{modeling}.  Using a reference wavelength of 250 $\mu$m, this opacity index corresponds to an opacity $\kappa$ of 0.19 cm$^2$g$^{-1}$ at 1.3 mm. Assuming a dust temperature of 30 K, and an optically thin system, our measured flux at 1.3 mm corresponds to a mass of 220 M$_{\odot}$ and a beam averaged column density of 1.1$\times$10$^{25}$ cm$^{-2}$.  The mass and column density would of course be lower if we use different values for $\beta_{\rm op}$, but to maintain consistency in the comparison with T Tauri stars, we use $\beta_{\rm op}$ of 2.35 for the modeling.  Using the Interstellar Medium value for $\beta_{\rm op}$ of 2.0 ($\kappa$=0.35 cm$^2$g$^{-1}$) \citep{hild1983}, the mass and column density then would be 120 M$_{\odot}$ and 6.4 $\times$10$^{24}$ cm$^{-2}$.  With a value for $\beta_{\rm op}$ of 1.4 ($\kappa$=0.93 cm$^2$g$^{-1}$) \citep{osse1994}, the mass and column density become 45 M$_{\odot}$ and 2.4 $\times$10$^{24}$ cm$^{-2}$. These masses are, of course, affected by missing flux in the interferometer data.  Comparing our interferometric 1.3 mm flux to the single dish peak
 flux at this wavelength within the 11$\arcsec$ beam of the MAMBO bolometer obtained by \citet{beut2002hmpo}, we estimate that 
 less than 60\% of the flux within this 11$\arcsec$ beam is filtered out.


\subsection{Line Data}\label{linedata}

\begin{table*} 
\caption{Outflow properties derived from CO observations.  Total outflow mass M$_{\rm{t}}$, momentum $\rm{p}$, energy E, size, outflow dynamical age $\rm{t}$, outflow rate $\dot{\rm{M}}_{\rm{out}}$, mechanical force F$_{\rm{m}}$ and mechanical luminosity L$_{\rm{m}}$ are given.  Inputs are discussed in the text. \label{output}}
\begin{center}
\begin{tabular}{cccccccc}
\tableline
\tableline
M$_{\rm{t}}$ [M$_{\odot}$] & p [M$_{\odot}$ km/s] & E [erg] & size [pc] & t [yr] & $\dot{\rm{M}}_{\rm{out}}$ [M$_{\odot}$/yr] & F$_{\rm{m}}$ [M$_{\odot}$ km/s/yr] & L$_{\rm{m}}$ [L$_{\odot}$] \\
\tableline
12 & 190 & 3.1 $\times$ 10$^{46}$ & 0.42 & 32000 & 3.8 $\times$ 10$^{-4}$ & 6.0 $\times$ 10$^{-3}$ & 8.0 \\
\tableline
\end{tabular}
\end{center}
\end{table*}

Toward the continuum peak, there are only four spectral lines present in the 4 GHz bandwidth of the SMA data (see Fig.~\ref{spectra}).  Based solely on chemical evolution, it is difficult to determine whether this source has not yet reached the hot core phase, or whether it has already passed through this evolutionary phase.  Three of the four lines present toward the continuum peak, $^{12}$CO(2--1), $^{13}$CO(2--1), and SO(6--5), are not high density tracers, and C$^{18}$O(2--1)  appears to be affected by the outflow (see Sect.~\ref{rotation}). Hence, with the data at hand we are unable to infer the kinematics of the elongated structure detected in the continuum and discussed below.  Methanol, a molecule linked to rotation in \citet{fall2009}, is present a few ($\sim$7) arcseconds to the northeast of the primary peak but is extremely weak directly at the location of the primary source.

\subsubsection{Outflow}
Previous single dish observations by \citet{beut2002outflow} were taken during non-ideal weather conditions, so the quality of our current data set is significantly improved.  Nevertheless, the $^{12}$CO (2--1) emission is consistent with the outflow suggested by this older data set.  The SMA data in Fig.~\ref{outflow} demonstrate that our observations follow the morphology of the H$_{2}$ jets seen by \citet{davi2004}.  In both sets of observations, the two outflows observed are highly collimated.  The two outflows we detect are roughly perpendicular to one another and appear to emanate from two separate sources detected in our 1.3 mm dust continuum.  
Outflows associated with low mass star formation depend on the presence of a disk. While massive outflows are likely also gas entrained by jets driven by magnetohydrodynamic acceleration from a disk, a description of other methods is given in \citet{arce2007}.

Several instances of anomalous behavior are apparent in Fig.~\ref{outflow}.  First, in the southeast-northwest outflow associated with the primary continuum source, there is a blueshifted component on the northwest side which is otherwise associated with redshifted emission (approximately located at $\Delta$R.A.=4$\arcsec$, $\Delta$Dec=7$\arcsec$).  This might be an indication that the outflow has a small inclination angle relative to the plane of the sky such that we observe both receding and approaching sections of the outflow.  Next, it is interesting to note that both sides of the outflow associated with the secondary continuum source are dominated by blueshifted emission.  This region is dominated by redshifted emission in the single dish data (Fig.~\ref{outflow}) indicating that it is large scale emission and is filtered out in the interferometric data. 

As discussed in Sect.~\ref{30m}, we do not combine the single dish data with the interferometer data.  
Instead, we use the interferometer data to accurately determine the outflow morphology as described above, but rely on the single dish data to measure fluxes.  
We use the single dish fluxes to calculate the outflow mass, dynamical age and outflow energetics following the approach of \citet{cabr1990}.
Using an average spatial extent of 30$\arcsec$ for each $^{12}$CO outflow lobe, and maximum blueshifted and redshifted velocities of 10 and 17 kms$^{-1}$ respectively, 
we derive the properties listed in Table~\ref{output}. Our derived values 
match those of \citet{beut2002outflow} reasonably well, and are likely better estimates due to poor weather conditions at the time of the earlier observations. These values confirm the original claim that the region will eventually form a high-mass star.

\subsection{Spectral Energy Distribution}\label{sedtext}
\begin{table} 
\caption{Sources of SED data points. \label{seddata}}
\begin{tabular}{ccccc}
\tableline
\tableline
$\lambda$ & flux$\footnote{The 450 and 800 $\mu$m fluxes are a factor of 4 smaller than the single dish fluxes published in \citet{mars2008} in order to make a more direct comparison with the interferometer data.  Further details are described in Sect.~\ref{sedtext}.}$ & +/- & aperture & reference \\
($\mu$m)  & (Jy) & (Jy)  &  ($\arcsec$)     &    \\
\tableline
12.8 & 26.3 & 2   & 5.0 & H. Linz, \emph{priv.~comm.}\\
18.7 & 41.0 & 2   & 5.0 & H. Linz, \emph{priv.~comm.}\\
24.8 & 101  & 30  & 1.9 & \citet{camp2008} \\
450  & 12.4 & 3.7 &   & \citet{mars2008} \\
800  & 1.9  & 0.6 &   & \citet{mars2008} \\
1300 & 0.4  & 0.1 & 3.3 & these data \\
\tableline

\end{tabular}
\end{table}
For the purpose of modeling, we gathered data for the spectral energy distribution (SED) of IRAS 18151-1208.
The wavelengths and associated fluxes that we include in our SED (Fig.~\ref{SED}) are tabulated in Table~\ref{seddata}.  The 1.3 mm data point for the SED is determined by measuring the interferometric flux of IRAS 18151-1208 within the central 10,000 AU by 10,000 AU region.  We note that the flux measured with the interferometer is a factor of four smaller than the corresponding  integrated single dish flux reported by \citet{mars2008}. The discrepancy between a factor 4 here and a factor of about 2 discussed at the end of Sect.~\ref{mmcontinuum} comes from the fact that \citet{mars2008} integrate over a much larger area than just the 11'' beam of the MAMBO observations. Since we would also like to use the 450 and 800 $\mu$m fluxes reported in \citet{mars2008}, we also divide these fluxes by a factor 4 since they were measured over the same spatial scales as the 1.2 mm data.
 We justify this approach by the fact that the SED approximately follows a power law in this wavelength regime.


For the SED, the 24.8 $\mu$m point comes from \citet{camp2008} and the 12.8 $\mu$m and 18.7 $\mu$m data points come from VISIR data (ESO proposal 075.C-0454, PI Fuller).  The VISIR data were taken on 5 August 2006 with the VLT.  The observations were made with the Q2 and [Ne{\sc ii}] filters which have central wavelengths of 18.72 $\mu$m and 12.81 $\mu$m and widths of 0.88 $\mu$m and 0.21 $\mu$m respectively.  Integration time with the Q2 filter was 6 minutes and 5.5 minutes with the [Ne{\sc ii}] filter. The standard star HD169916 was used for
flux calibration \citep{cohe1999}.  

The nominal peak positions vary by $\sim$1$\arcsec$ between the millimeter and mid-infrared data.  However, in the mid-infrared, a small field of view does not allow high positional accuracy so we attribute the emission to the same source.  Alternatively the offset may be due to a real difference in spatial position which would lead to a lowering of the SED at these wavelengths.  We note this possibility, but proceed assuming that the detections at different wavelengths are spatially coincident.

The data of \citet{camp2008} exhibit a silicate absorption dip at 10 $\mu$m.  
This dip is likely caused by absorption of the envelope.  Since  the millimeter interferometer data filter out the envelope, we are primarily interested in the small scale disk-like structure, so we do not try to reproduce this feature.

\section{Modeling} \label{modeling}
After the 3D radiative transfer modeling of the circumstellar environment in the low mass regime was proven successful \citep{wolf2003,saut2009}, we wanted to test whether a similar method could work in the high mass regime.  We hypothesized that a scaled up version of a T Tauri disk may be able to reproduce the observations of what may possibly be a disk around a massive protostellar object.

 While we are aware that the disk in IRAS 18151-1208 is still embedded in  a large-scale envelope, 
we prefer to model only the disk structure. 
The interferometric map clearly shows an elongated structure, and while it may still have some contribution from a flattened envelope, 
applying a disk model to this structure appears to be a reasonable approach. Furthermore, while we have missing flux on scales of the SMA primary beam (55$\arcsec$) on the order of 75\%, it is  only on the order of 60\% when going to scales of the 11$\arcsec$ beam of the IRAM 30m observations (Sect.~\ref{mmcontinuum} and \ref{sedtext}). Going to scales of the SMA synthesized beam (0.8$\arcsec$x0.7$\arcsec$), the missing flux is even less severe. Assuming that the single-dish peak flux within the 11$\arcsec$ beam of 980 mJy (Beuther \emph{priv.~comm.}) is smoothly distributed, the single-dish flux per SMA beam is then 980 mJy/[11$\arcsec$x11$\arcsec$/(0.8$\arcsec$x0.7$\arcsec$)] $\sim$ 3 mJy which is only 1.5 times the rms of the SMA continuum image (see Tab.~\ref{obs_params}). Furthermore, we can estimate the extinction at mid-infrared wavelengths due to the envelope.  Based on an extinction A$_{V}$ of 72 mag \citep{camp2008}, assuming R$_{V}$ = 5.0 we estimate the extinction at 12 $\mu$m to be approximately 2 mag \citep{math1990}.  Therefore, although a component is missing,  modeling only the disk appears to be a reasonable approximation for comparison purposes with low-mass disk models.

\subsection{The Monte Carlo 3D Radiative Transfer Code}
Using the self-consistent 
3D Monte Carlo Radiative Transfer code, MC3D, developed by \citet{wolf1999,wolf2003mc3d}, 
we model the elongation seen in the dust continuum of IRAS 18151-1208.  
In our setup, MC3D calculates the density distribution in a spherical region surrounding the central star 
divided into 91 equally spaced angular divisions and 50 logarithmically spaced radial divisions such that the successive grid cell is $\sim$1\% longer than its inner neighbor.  This distribution is chosen to better resolve the density gradient 
in the more massive 
region closer to the central star.  
Using 100 logarithmically distributed wavelengths between 0.05 and 2000 $\mu$m, MC3D computes the thermal structure of the disk, derives SEDs, and produces images at any of the 100 wavelengths. For the purpose of this work, we use all 100 wavelengths for the temperature structure and the SED, and we compute an image at 1.3 mm.

\subsection{The Disk}
We use a density profile of a parameterized rotationally symmetric disk 
to produce fits to the observed SED and 
brightness profile along the elongation.  To maintain a more even comparison, we assume the density profile used in the low mass cases \citep{wolf2003,saut2009} although we cannot exclude other density profiles with this approach.   
The density distribution we use is described by the following equation:

\begin{equation} 
  \rho_\mathrm{disk}(r_\mathrm{cyl}) = \rho_{0}\left(\frac{R_{*}}{r_\mathrm{cyl}}\right)^{\alpha} \exp{\left[-\frac{1}{2}\left(\frac{z}{P_\mathrm{h}}\right)  ^2\right] }
\label{density}
\end{equation}
where $z$ is the height above or below the midplane, R$_{*}$ is the radius of the central star and $r_{cyl}$ is the radial distance from the central rotation (z-)axis.  The 
normalization constant, $\rho_0$, is determined by the conditions placed on the total mass and volume of the system, and
\begin{equation}
 P_h=h_{100}\left(\frac{r_\mathrm{cyl}}{100 \mathrm{AU}}\right)^\beta
\label{scale_height}
\end{equation}
where $h_{100}$ is the scale height of the disk 100 AU from the central star.  The shape of the modeled disk is described by the parameters $\alpha$ and $\beta$ which are the radial density parameter and the disk flaring parameter respectively.  If $\alpha$ and $\beta$ satisfy the relation $\alpha$ = 3($\beta$-$\frac{1}{2}$), then the density distribution corresponds to that of \citet{shak1973}. The dependence of the surface density on the disk shape parameters can then be obtained by integrating Equation \ref{density} with respect to $z$:
$\Sigma$($r_\mathrm{cyl}$) $\sim r_\mathrm{cyl}^{p}$ where p=$\beta$-$\alpha$.

\subsection{The Dust Model}\label{dustmodel}
Under the assumption that the gas is optically thin in the wavelength regime we are working in, it is the dust that is described by the density distribution in Eq. \ref{density}.  The dust grains are assumed to be spherical, and we adopt a power law size distribution following the MRN \citep{math1977} approach where the number of dust grains of a given radius $a$ is given by 
\begin{equation}
 n(a) \sim a^{-3.5}.
\end{equation}
Dust grain sizes $a$ vary from 5 nm to 250 nm which are typical of the interstellar medium.

We tested two different chemical compositions.  The first follows the graphite and smoothed silicate mixture of \citet{wein2001} with relative abundances of 37.5\% (graphite) and 62.5\% (astronomical silicate). The other contains by mass 24.2\% iron and iron-poor silicates, 5.6\% troilite, 25.6\% refractory organics, 4.4\% volatile organics, and 40.3\% water \citep{poll1994}. 

\subsection{The Parameter Space} \label{params}
A motivating question behind this study is whether a low mass disk can be scaled up to describe a high mass disk.  This would manifest itself in the form of having similar parameters describing the shape of the disk and scaled up parameters for the quantities such as disk mass and extent, accompanied by a more massive central star. In the scenario that scaling-up reasonably describes the observations of the high mass protostellar object in question, we would expect the values for $\alpha$ and $\beta$ to remain the same as in the low mass case.  Although the value for h$_{100}$ between the two cases is not directly comparable because 100 AU represents very different regions of the respective disks, we can scale h$_{100}$ to a proportionally similar fraction of the disk's size and expect that to be proportionally larger than in the low mass case.  In the ideal case, the disk mass should scale relative to the volume of the system which can be varied 
via the outer radius.

The parameters that we include in the modeling are:
\begin{itemize}
\item $\alpha$, $\beta$, and h$_{100}$: the density distribution exponent, the disk flaring exponent, and the scale height of the disk at 100 AU respectively.  These quantities describe the dust density distribution via Equations \ref{density} and \ref{scale_height}.  We 
test 13 values of $\alpha$ between 1 and 4 and 11 values of $\beta$ between 0.5 and 3.  The relation $\alpha$ = 3($\beta$-$\frac{1}{2}$) is shown to hold in the case of low mass disks \citep{wolf2003,saut2009}.  We originally varied $\alpha$ and $\beta$ independently.  However, after discovering that modifying $\alpha$ alone did not cause significant changes in the modeling results, subsequent models followed the $\alpha$ = 3($\beta$-$\frac{1}{2}$) relation. 
We test 8 values of h$_{100}$ between 1 and 30 AU.
\item $\theta$: the inclination of the disk with respect to the plane of the sky.  The observations indicate the presence of a flattened object, and this would only be visible at inclination angles close to edge-on.  However, we test 6 inclination angles ranging from 45$^{\circ}$ to 90$^{\circ}$ (edge-on).
\item R$_\mathrm{in}$: the inner radius.  We initially chose small values for R$_\mathrm{in}$ of 10, 20, and 50 AU, but also tried significantly larger values such as 400, 1000, 2500, and 4000 AU before settling on 20 AU in our best fit model.
\item R$_{\rm{out}}$: the radius at which the density goes to zero.  The outer radius can roughly be constrained by the 1.3 mm continuum observations to a nominal value of 5,000 AU, so we assume this value throughout.  
\item M$_\mathrm{disk}$: the dust mass.  Using the dust model of \citet{wein2001} we determine a rough estimate for the mass within the 5,000 AU radius to be 220 M$_{\odot}$ (See details in Sect.~\ref{mmcontinuum}).  From this, we derive a dust mass of 2.2 M$_{\odot}$ based on a gas-to-dust ratio of 100.  Taking this as our starting estimate, we subsequently determine the mass by ensuring that the SED fit goes through the 1.3 mm data point.
\item Central star: due to the luminosity constraint of 20,000 L$_{\odot}$ described in the next section (Sect.~\ref{constraints}), the most massive star we consider is a B0.5[T=26300 K, L=20000 L$_{\odot}$, R=7.0 R$_{\odot}$, M=15 M$_{\odot}$].  We also test a B1[T=25400 K, L=16000 L$_{\odot}$, R=6.4 R$_{\odot}$, M=13 M$_{\odot}$] and a B2 [T=22500 K, L=6000 L$_{\odot}$, R=5.6 R$_{\odot}$, M=10 M$_{\odot}$] central star.
\end{itemize}

\subsection{Observational Constraints} \label{constraints}

For the SED, the fitting is constrained in a straightforward manner by the observed fluxes.  We attempt to fit our SED over the entire wavelength regime for which we have data points available.

While we use the SED to guide our efforts since the effects of modifying model parameters are more noticeable there (see Figs.~\ref{errorSED} and \ref{errorIMG}) than in the image, we aim 
to reproduce the elongated structure seen in the 1.3 mm dust continuum perpendicular to the outflow orientation. Among other interpretations of what may be taking place in the region, possible explanations include a double source and a scaled up low mass disk.  Here we focus on whether a scaled up low mass disk is consistent with these observations. 
Although we were able to reproduce a symmetric double peaked structure prior to smoothing with the PSF by making the inner radius larger, the convolution with the observed beam size smoothed out this effect in the image brightness profile.  We then shifted our focus to fitting the shape of the elongated structure's outer edges.  This is not a significant oversight because the difference in the height of the two peaks is only on the order of 1$\sigma$.  It is worth mentioning that using an extremely large inner radius flattens out the peak of the 
brightness profile along the major axis (see Fig.~\ref{errorIMG}), but because we do not know the origin of this slight asymmetry, we do not include anything in the modeling to account for the specific asymmetry nor the double peak and therefore do not give it any extra weight in our determination of the best fit model. 

For the model parameters, we were able to place rough constraints on the central star, the outer radius, and the mass of the system before modeling began.  \citet{srid2002} place the luminosity of the IRAS 18151-1208 region at 20,000 L$_{\odot}$.  We therefore use this as a limiting luminosity for the central star in our modeling.  We use an outer radius of 5,000 AU which is the radius at which the midplane density goes to zero and corresponds approximately to the 4$\sigma$ contour of the elongation in the 1.3 mm dust continuum.  
Making the assumption that the region is optically thin, we can obtain a starting estimate for the mass of the system following the approach described in Section \ref{mmcontinuum}.  Our measured flux at 1.3 mm corresponds to a mass of 220 M$_{\odot}$.  We take this as a starting estimate, but since there are many sources of uncertainty in this calculation such as missing flux and unknown optical depth to name a few, we 
did not strictly limit the mass in our tests to this value.
For the test cases with a gas mass of 220 M$_{\odot}$, slight differences on the order of a few percent between the observed flux and the calculated flux at 1.3 mm sometimes come about. 
We attribute this to the fact that despite using the same dust model, MC3D takes optical depth into consideration while the mass estimate does not.

\subsection{Comparing the Model with Observations} \label{comp_mod_obs}
\begin{figure}
\includegraphics[scale=0.5]{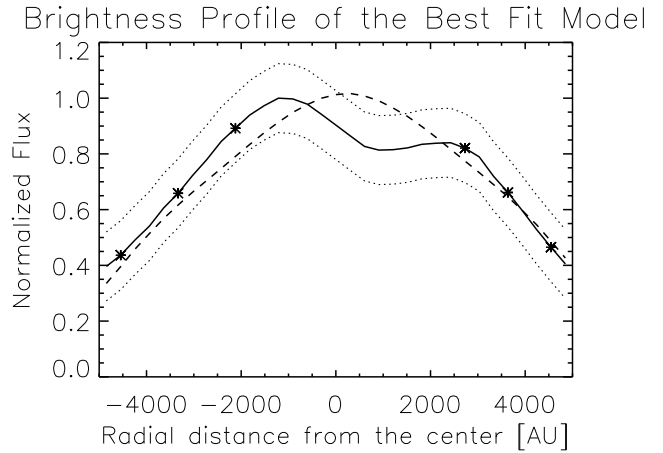}
\caption{The 
density profile along the elongation axis of the best fit model. The solid line is the cut through the midplane of the observed elongated structure.  The dotted lines are +/- 1$\sigma$, and the dashed line is the model. The data points used in the weighted least squares fit are included in the plot. \label{rad_den}}
\end{figure}

\begin{figure}
\includegraphics[scale=0.5]{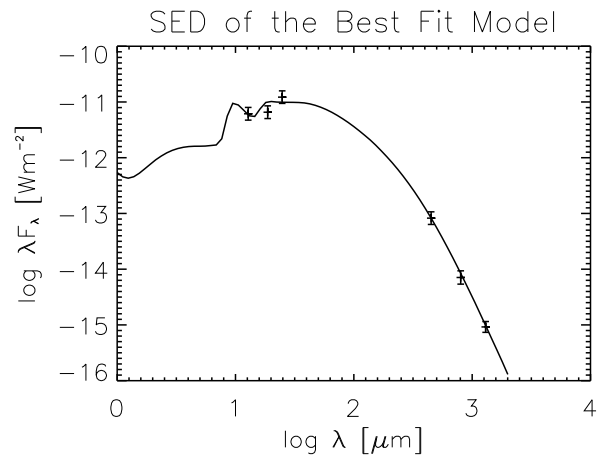}
%
\caption{The SED of the best fit model.  Due to differences in the observation methods, the 450 and 800 micron data points have been adjusted for a more direct comparison with the 1.3 mm data point (see Sect.~\ref{sedtext} for details.) 
 \label{SED}}
\end{figure}

The parameters of the best fit model are summarized in Table~\ref{modelparams} and for comparison, the best fit model of the Butterfly Star \citep{wolf2003, wolf2008} and CB 26 \citep{saut2009} are also included in this table.  The corresponding SED and brightness profile of the best fit model are shown in Figs.~\ref{rad_den} and \ref{SED}.  All data points in the SED as well as the six points shown in Fig.~\ref{rad_den} were used to determine the best fit model.

We determine the best fit model based on considering both the SED and the 
brightness profile of the 1.3 mm image.  For each test we calculate a reduced chi-squared value as described in the last paragraph of the next section (Sect.~\ref{testing_params}). 
The modeled image is convolved with the observed PSF and the 
brightness profile along the major axis is then obtained by summing the flux from the pixels in the columns perpendicular to the elongation.  This is shown as the dashed line in Figs.~\ref{rad_den} and \ref{errorIMG}.  The pixels from columns aligned perpendicular to the elongation in the observed image are also summed and this is plotted as the solid line in Figs.~\ref{rad_den} and \ref{errorIMG}.  A comparison between the observed and modeled brightness curves is made by calculating the difference between the two curves at the six points indicated in Fig.~\ref{rad_den}.  Weighting each point in the image and SED equally, we combine these differences with the difference between the six observed and modeled SED points to arrive at the final value used to determine the best fit model.  The numerical value of the weighted least squares calculation cannot be used on its own, but compared to the values from other models, we have a numerical value on which to base our decision of the best fit. Our best fit had a chi square value of 2.4.  While many models had chi square values less than 10, some were as high as 30.

\begin{table*}
\caption{The parameters used in the best fit model. \label{modelparams}}
\begin{center}
\begin{tabular}{ccccccccccc}
\tableline
\tableline
Object & $\alpha$ & $\beta$ & h$_{0}$ & M$_\mathrm{dust}$ & $\theta$ & R$_\mathrm{in}$ & R$_{\rm{out}}$ & R$_{*}$ & T$_{*}$ & L$_{*}$ \\
   &  &  & (AU) & (M$_{\odot}$) & ($^{\circ}$) & (AU) & (AU) & (R$_{\odot}$) & (T$_{\odot}$) & (L$_{\odot}$) \\
\tableline
IRAS 18151\tablenotemark{a} & 2.4  & 1.3  & 200  & 2.2    & 60 & 20 & 5000 & 6.4 & 25400 & 16000 \\
IRAS 04302\tablenotemark{b} & 2.37 & 1.29 & 15 & 0.0007 & 90 & 0.07 & 300  & 2   & 4000  & 0.92 \\
CB 26\tablenotemark{c}      & 2.2  & 1.4  & 6 & 0.003  & 85 & 45   & 200  & 2   & 4000  & 0.92 \\
\tableline
\end{tabular}
\end{center}
\footnotesize{
~\\
Parameters include the radial 
density distribution parameter $\alpha$, flaring parameter $\beta$, scale height of the disk at $\frac{1}{3}$R$_\mathrm{out}$ h$_{0}$, dust mass M$_\mathrm{disk}$, inclination angle $\theta$ ($\theta$=90$^{\circ}$ corresponds to an edge-on disk), the inner and outer radii of the density distribution R$_\mathrm{in}$ and R$_{\rm{out}}$ and the radius, temperature, and luminosity of the central star.}
\tablenotetext{a}{This work}
\tablenotetext{b}{\citet{wolf2003}}
\tablenotetext{c}{\citet{saut2009}}
\end{table*}

\section{Discussion}
\begin{figure*}
\subfigure{\includegraphics[width=0.5\textwidth]{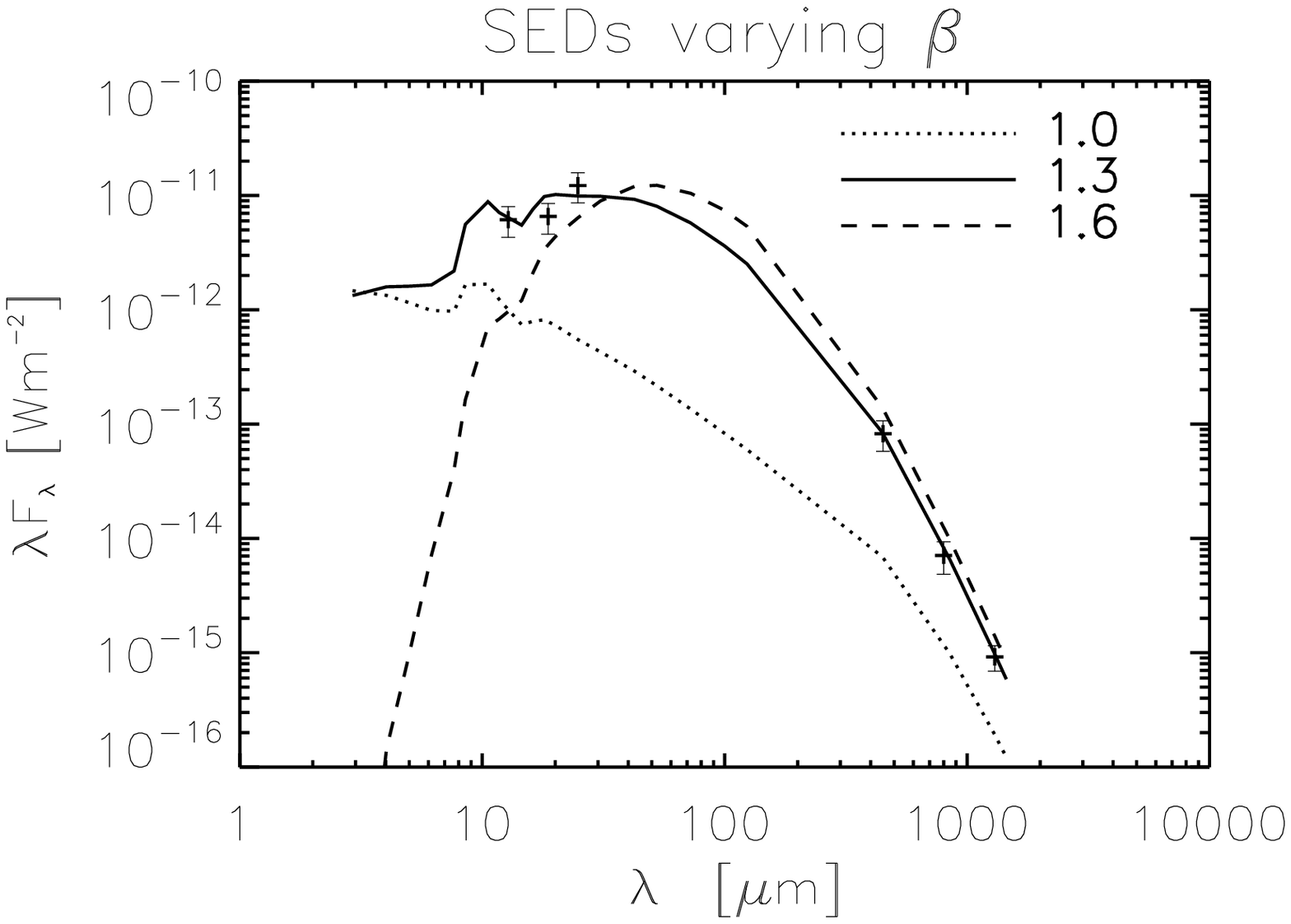}}
\subfigure{\includegraphics[width=0.5\textwidth]{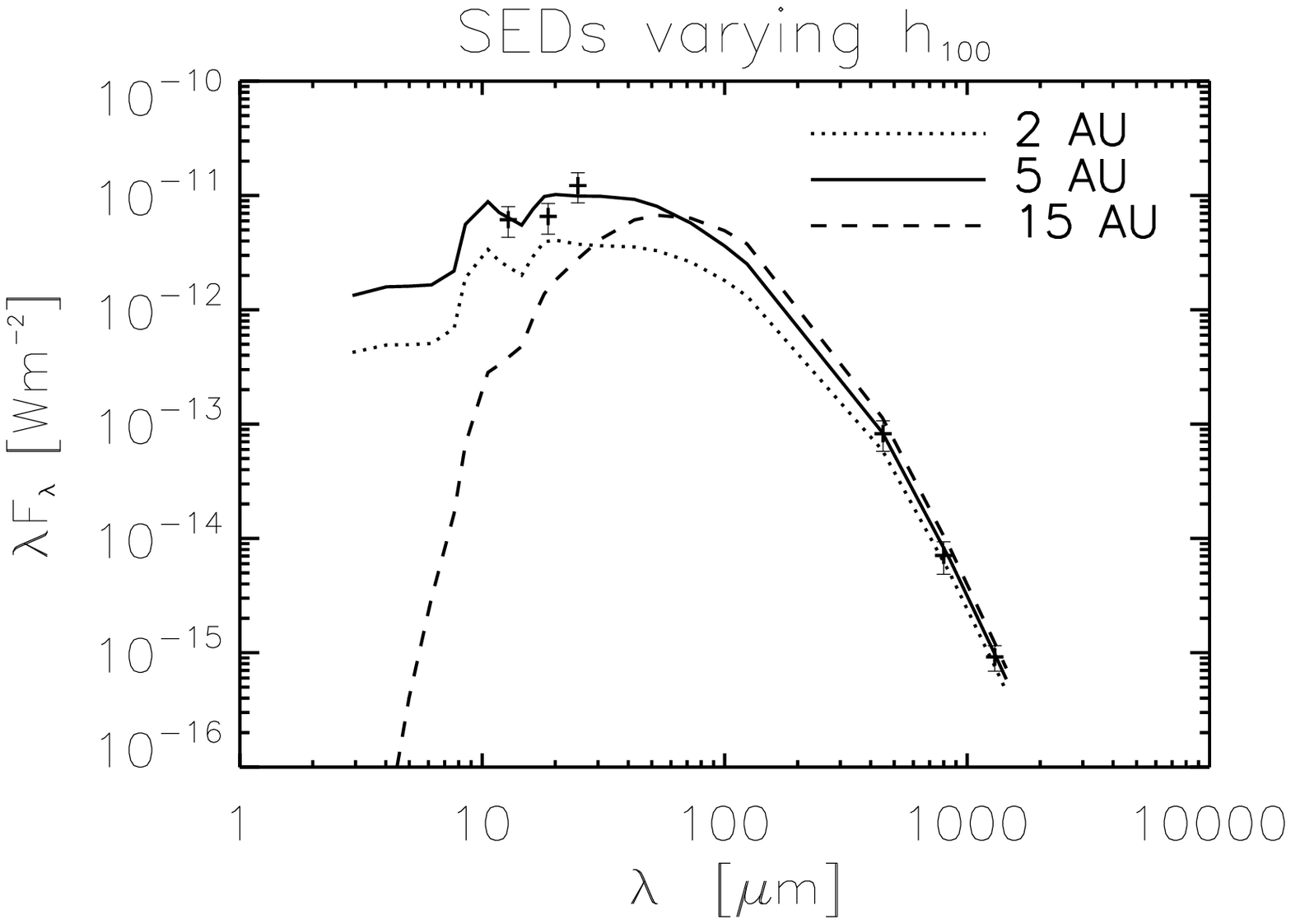}}
\subfigure{\includegraphics[width=0.5\textwidth]{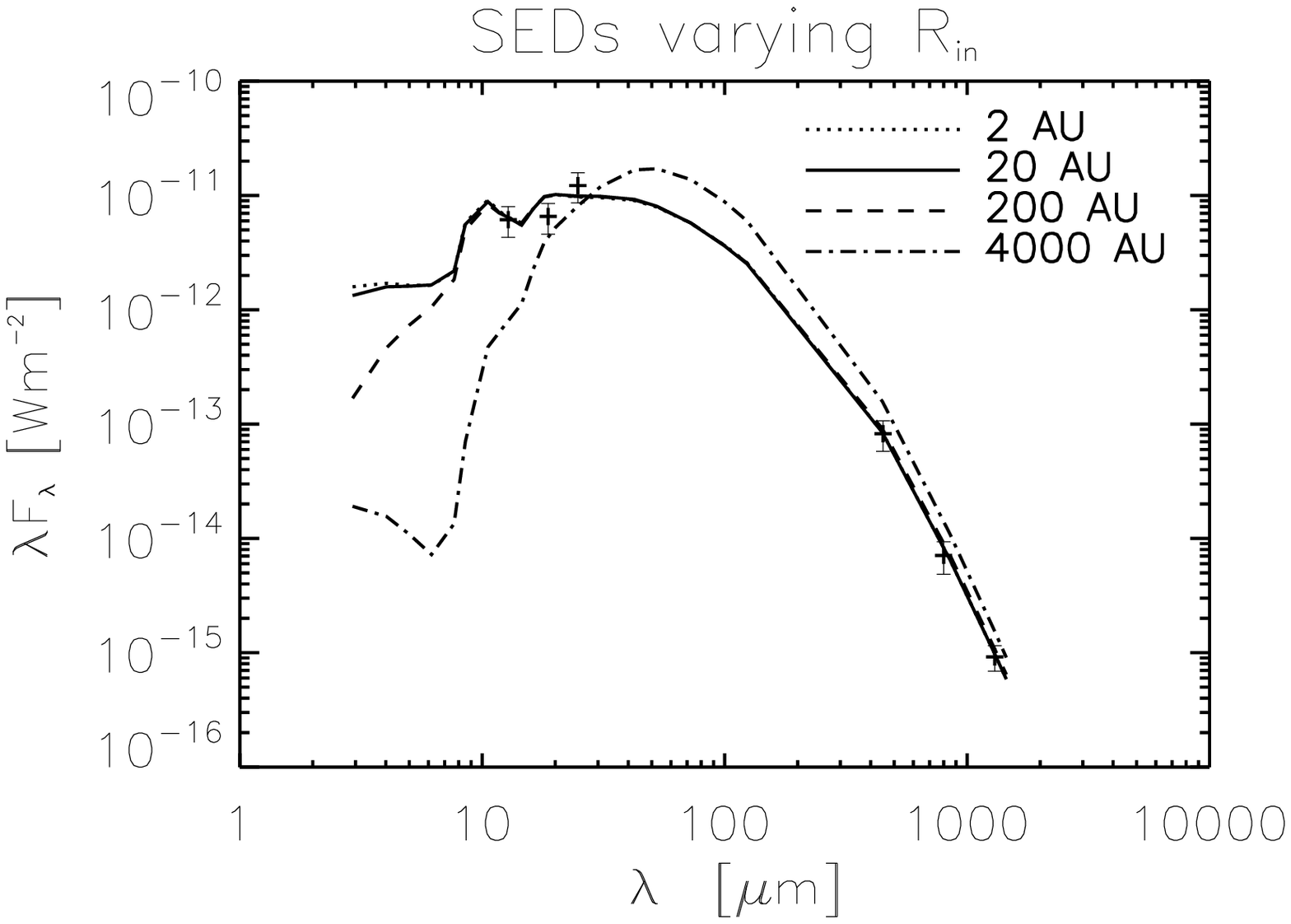}}
\subfigure{\includegraphics[width=0.5\textwidth]{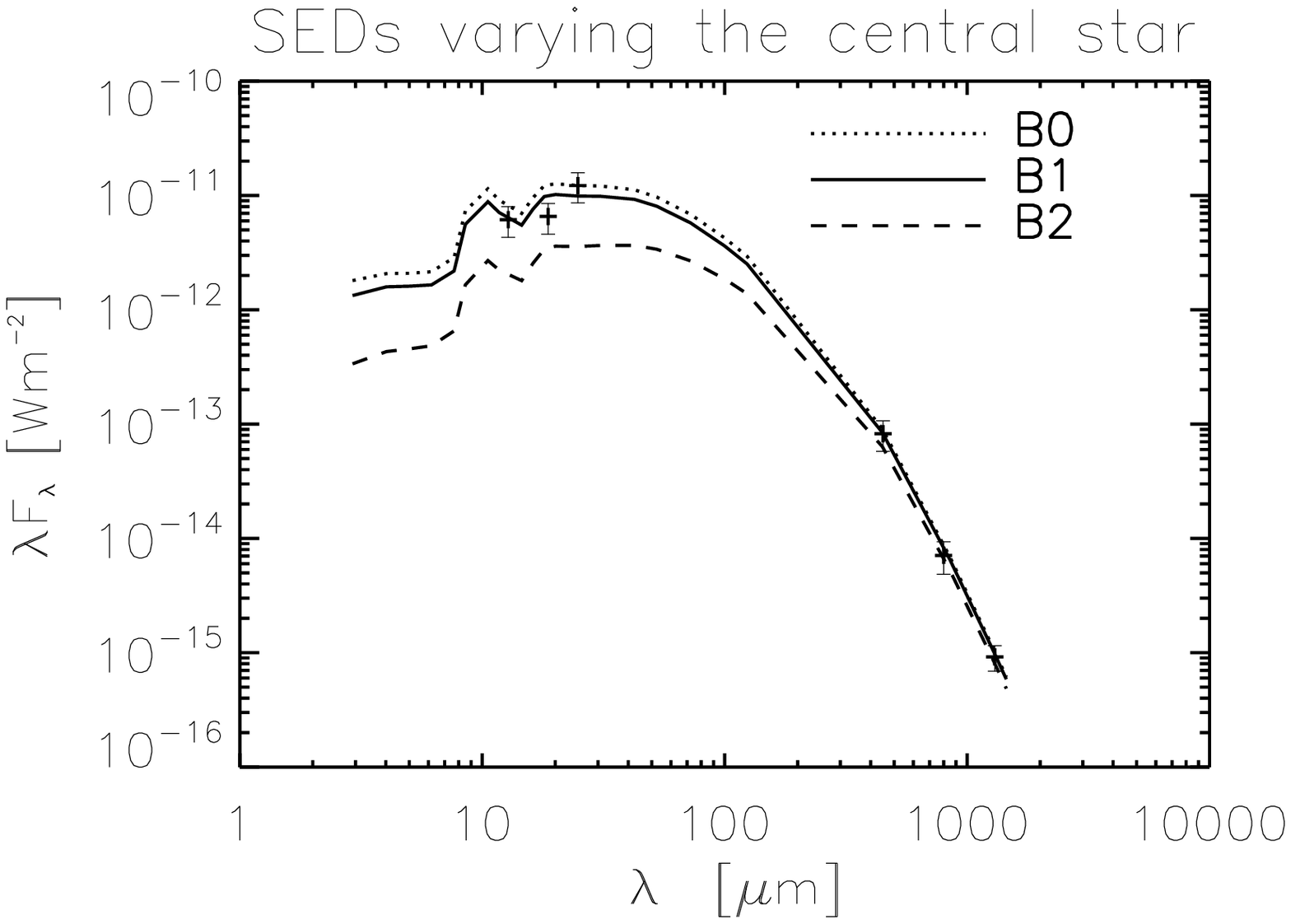}}
\subfigure{\includegraphics[width=0.5\textwidth]{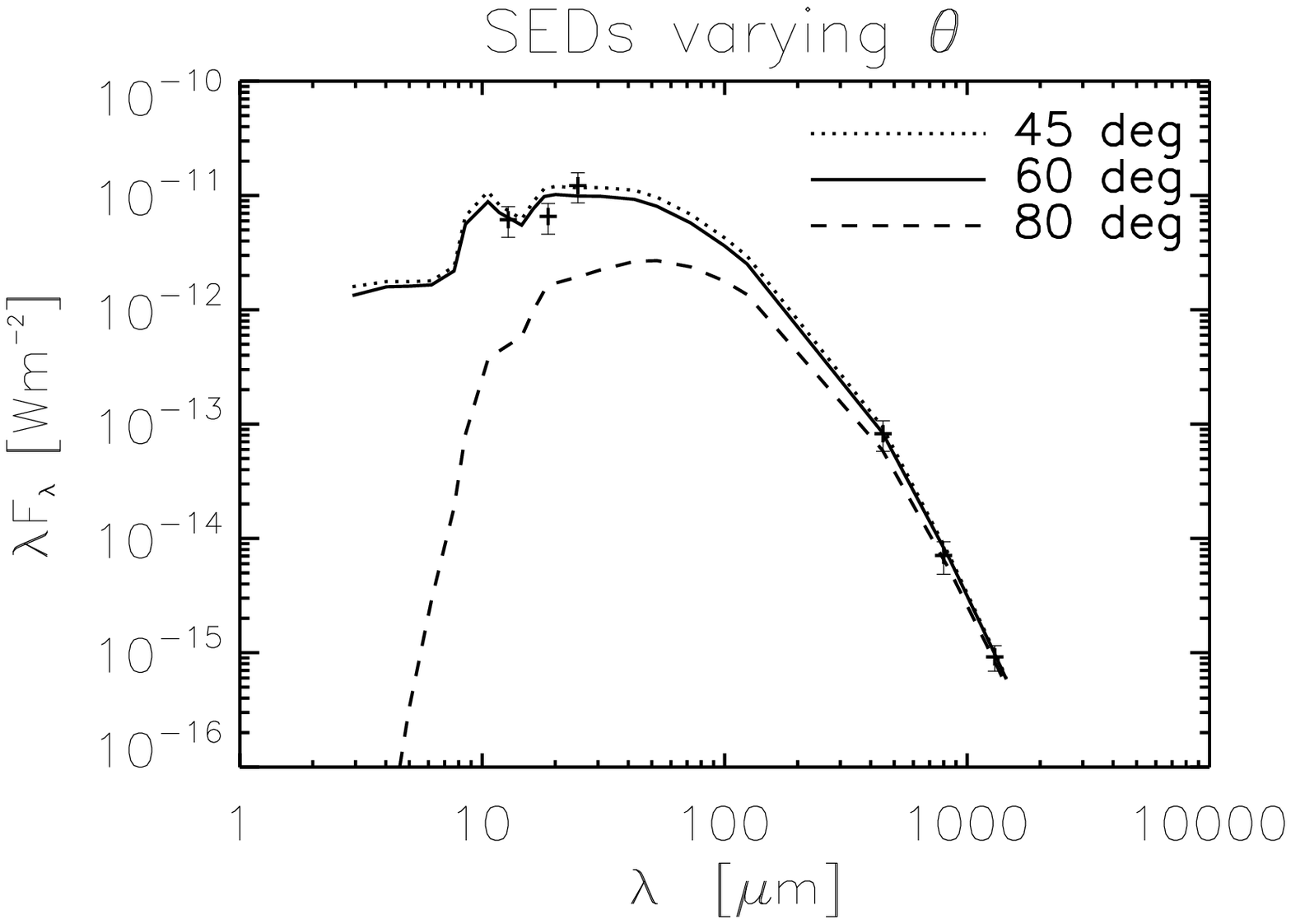}}
\subfigure{\includegraphics[width=0.5\textwidth]{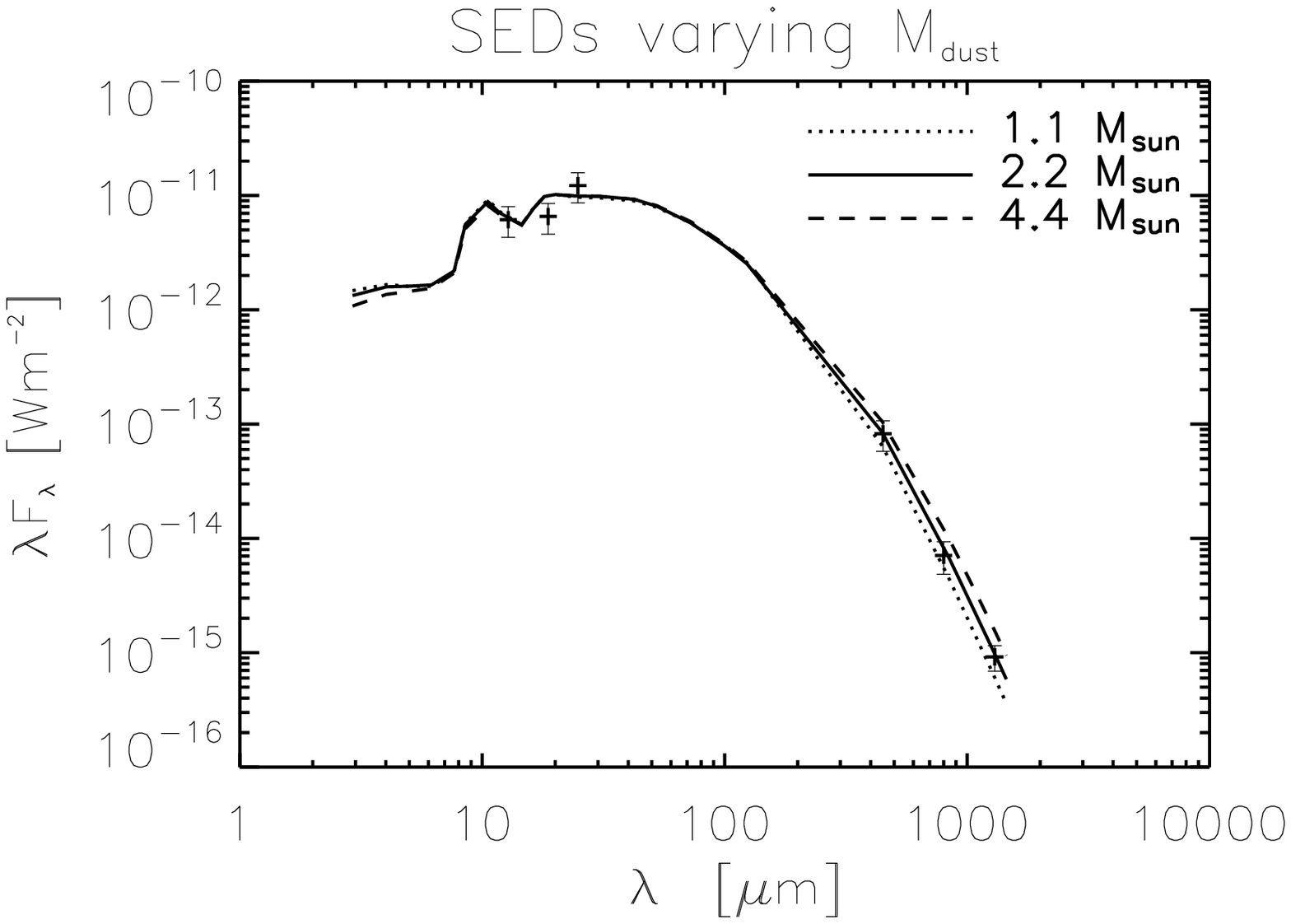}}

\caption{The effect on the SED of varying individual parameters.  Although numerous combinations of parameters were tested (see Sect.~\ref{params} for a description of the parameter space), these demonstrate how individual parameters affect the SED.  The solid line is always the best fit model (see Fig.~\ref{SED}). SEDs varying (from left to right, top to bottom) the 
flaring exponent ($\beta$), the scale height at 100 AU (h$_{100}$), the inner radius (R$_\mathrm{in}$), the central star's spectral type, the inclination angle ($\theta$), and the dust mass (M$_\mathrm{disk}$). \label{errorSED}}
\end{figure*}

\subsection{Testing the Parameter Space}\label{testing_params}
\begin{figure*}
\includegraphics[scale=1]{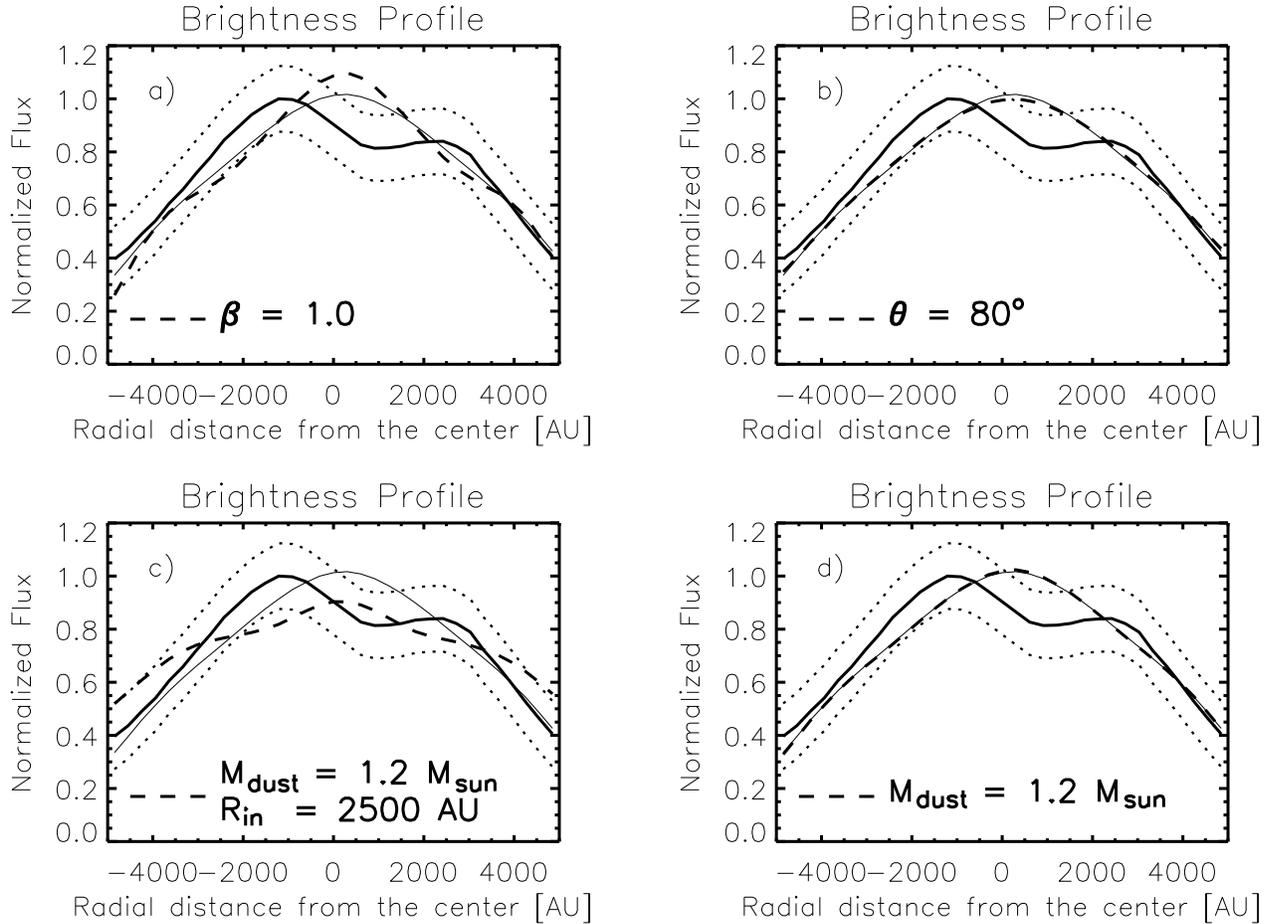}

\caption{The effect on the 1.3 mm image's 
brightness profile of varying individual parameters. In each panel, the thick solid line is the cut through the midplane of the observed elongated structure, the dotted lines are +/- 1$\sigma$, the dashed line is the model and the thin solid line is an overlay of the best fit model (see Fig.~\ref{rad_den}). Here we have included only a select sample of the most deviant cases.  The brightness profile was not affected as heavily as the SED and many of the cases show very minor deviations from the best fit model, so we do not include them here.  In choosing the best fit model, we took both the SED and the brightness profile into account.  See Sect.~\ref{comp_mod_obs} for details. The parameters are similar to those of the best fit model except for the parameter(s) indicated in each panel. In panel \emph{(a)}: 
the disk flaring parameter $\beta$=1.0 instead of 1.3.
\emph{(b)}: 
the inclination angle $\theta$=80$^{\circ}$ instead of 60$^{\circ}$
\emph{(c)}: 
the \textbf{inner} radius R$_{\rm{in}}$=2500 AU instead of 20 AU and the dust mass M$_\mathrm{dust}$=1.2 M$_{\odot}$ instead of 2.2 M$_{\odot}$.
\emph{(d)}: 
the dust mass M$_\mathrm{dust}$=1.2 M$_{\odot}$ instead of 2.2 M$_{\odot}$.\label{errorIMG}}
\end{figure*}

As described in Sect.~\ref{params}, a wide range of model parameters were tested before settling on the ones used in the best fit model.  Figures~\ref{errorSED} and \ref{errorIMG} contain overlaid SEDs and 
brightness profiles from along the elongation that demonstrate the effects of varying the model parameters.  The effects on the SEDs are conspicuous, but the differences in the brightness profiles are more subtle.  We therefore quantify each fit by determining a reduced chi-squared value for all tests performed.  Although we do not have enough information about the envelope component to make quantitative statements about its effects, we discuss here 
how its omission affects our modeling results.

As Figs.~\ref{errorSED} and \ref{errorIMG} demonstrate, some parameters can be constrained better than others.  We fix the outer radius, R$_{\rm{out}}$ based on the observations and although we still vary the values of the parameters for  the disk's dust mass m$_{\rm{dust}}$ and the luminosity, temperature and radius of the central star, 
 we do this mainly to see how varying these parameters affects the model.  Ultimately, the values for these parameters were determined by the 1.3 mm observations rather than through the modeling process.  By not including an envelope component in the model, all of the millimeter dust emission is attributed to the disk.  This means that the disk dust mass is likely overestimated. 
 The outer radius as defined by the millimeter image may also be overestimated if the actual disk component is being obscured by an envelope.  The inner radius, on the other hand, is not well constrained by our models, and the addition of an envelope component would not affect the situation.  

The flaring exponent $\beta$ is constrained relatively well across the entire infrared and submillimeter wavelength regime as well as by the brightness profile, and we choose the corresponding value for the density distribution exponent $\alpha$ based on the relation known for low mass disks \citep{shak1973}.  
Because some of the disk may be obscured by the envelope component and especially the less dense outer flared regions, our value for the amount of flaring, characterized by $\beta$, may be higher 
than if an envelope had been included in the modeling.   Because of the $\beta$ - $\alpha$ relation,  $\alpha$ would be similarly affected.  However, the actual nature of the 
overestimation depends on the density distribution in the envelope, e.g. its radial power-law.  

The parameters for the disk scale height h$_{100}$, and the inclination angle 
$\theta$ are predominantly constrained only by the SED in the mid-infrared wavelength regime.  However, the inclination angle is further limited by the millimeter image as it is seen as a flattened elongated object and hence is closer to edge-on opposed to face-on.  As the envelope component plays more of a role at these shorter wavelengths, the values derived for these parameters should be treated with some caution. In the presence of a thick spherically symmetric envelope, the elongation of the disk would still be observed, providing support for the limitation on the inclination angle, but other envelope geometries may have different effects on the observed component.  Using similar reasoning to the disk shape parameter $\beta$, the disk scale height would also be an  overestimate 
because of the envelope component obscuring the outer part of the disk.

Varying the dust composition as described in Section \ref{dustmodel} did not  
significantly affect our results, so we present here only the results of the \citet{wein2001} composition to maintain a more direct comparison with the Butterfly Star and CB 26.

\begin{figure}
\includegraphics[scale=0.5]{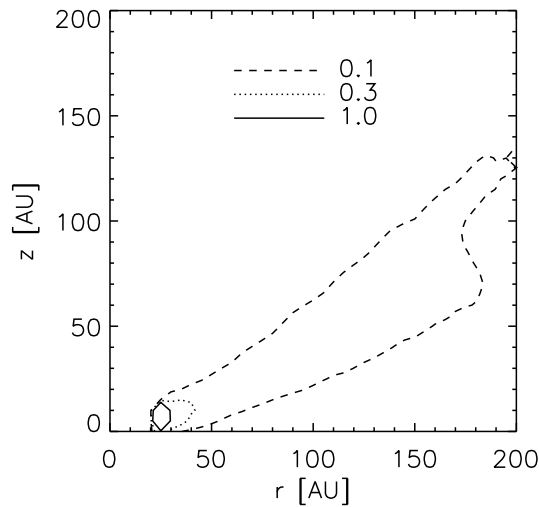}
\caption{Contour plot of the Toomre Q stability parameter. The value of Q is labeled next to the respective contour. Values of Q greater than 1.0 indicate stability under the Toomre criterion. \label{ToomreQ}}
\end{figure}

\begin{itemize}
\item The disk shape parameters 
$\beta$ (flaring parameter), and h$_{100}$ (scale height at 100 AU) strongly 
affect the shorter wavelengths in the SED.  
As demonstrated in Fig.~\ref{errorSED}, both lower and higher values of $\beta$ and of h$_{100}$ than used in the best fit model significantly decrease the flux in the shorter wavelength (micron) regime of the SED.  Although the SED may be affected by a complex interplay between several effects, it is likely that the optical depth is primarily responsible for this behavior.  
Smaller values of beta ($\beta <$ 1.0) mean that the disk is self-shadowed and thus less surface area is available for absorption and reemission of radiation. A more flared disk (larger value of $\beta$) obscures the central star more because of its larger surface area which intercepts more of the star's radiation. 
Both of these effects lead to a reduction of mid-infrared flux in the SED.  For h$_{100}$, thicker disks (larger values) hide the central star while smaller values have the same amount of mass distributed over a smaller volume leading to higher density which increases the optical depth.  As a result, the temperature gradient becomes steeper leaving a smaller area for reradiation which in turn means less flux in the micron regime. 
\item The density distribution exponent $\alpha$ cannot be constrained by the data themselves. Higher resolution data would be necessary to provide stronger constraints on $\alpha$.  After constraining $\beta$, we use the direct relation for \citet{shak1973} disks between $\alpha$ and $\beta$ [$\alpha$ = 3($\beta$-$\frac{1}{2}$)] to determine a value of 2.4 for $\alpha$.  This approach satisfactorily reproduces our observations. 
\item Inclination angle:  the difference in the SEDs is small for inclination angles between 45$^{\circ}$ and 60$^{\circ}$, but we use 60$^{\circ}$ in our best fit model based on the observational data that indicate a more edge-on orientation.  The SED deviates further from the observations at inclination angles larger than 60$^{\circ}$.  At very high inclination angles (e.g. 80$^{\circ}$), the upper layers of the disk block the central star from view and the SED drops significantly in the micron regime.
\item Disk mass: in the mass regime we tested, the only parameter that significantly affects the millimeter wavelength side of the SED without affecting the micron side is the dust mass. Figure~\ref{errorSED} makes it apparent that using the dust model of \citet{wein2001}, a dust mass of 2.2 M$_{\odot}$ is necessary to fit the observed 1.3 mm flux.  With a gas-to-dust ratio of 
\begin{equation}
\frac{M_\mathrm{Gas}}{M_\mathrm{Dust}}\sim100,
\end{equation}
this corresponds to a total disk mass of 220 M$_{\odot}$.  This is much larger than the main sequence mass of a B1 central star (13 M$_{\odot}$) indicating that this system cannot be Keplerian and should be quite unstable.  To quantify this statement, we calculate the stability in the system under the assumption that it is a disk.  Following \citet{toom1964}, the condition for stability is 
\begin{equation}
 Q(r,z) = \frac{c_\mathrm{s}(r,z)\kappa_\mathrm{k}(r)}{\pi G\Sigma(r)} > 1
\end{equation}
where c$_\mathrm{s}$ is the local sound speed, $\kappa_\mathrm{k}$ the angular frequency of the disk derived from Kepler's Law, G the gravitational constant and $\Sigma$ the surface density.  Figure~\ref{ToomreQ} indicates that 
regions only within the central-most 30 AU of the system are stable under the Toomre criterion.  
These results are consistent with what \citet{krat2006, vaid2009} find in their modeling. 
As expected for a star-disk system that has a significant fraction of the mass in the disk, the modeled disk in this situation is unstable and is likely fragmenting and possibly forming a multiple system as described by \citet{krum2009}.
\item Central star: because we do not have observational data for the wavelength regime dominated by the central star, we do not include these wavelengths in our SED modeling 
since the main contribution to the flux at the wavelengths for which we have data is from the warm dust in the disk.  Nevertheless, increasing the temperature and luminosity of the central star raises the SED in the dust dominated wavelength regime and shifts the peak to shorter wavelengths as one would expect for warmer objects. 
\item Inner radius:  we do not expect our symmetric model to reproduce the asymmetric peak structure of the observed 
brightness profile along the elongation but even having a large inner hole (large R$_\mathrm{in}$) does not produce a double peak structure as one might expect.  At extreme values of R$_\mathrm{in}$ such as 2500 or 4000 AU, the brightness profile does appear double peaked, but after convolution with the observed beam, the double peak that comes out of the modeling is smoothed out.  This is not surprising given the $\sim$2400 AU spatial resolution of the data (see Table \ref{obs_params}). Increasing the inner radius to these extreme values flattens out the brightness profile toward the center (see Fig.~\ref{errorIMG}) but also makes the brightness profile wider which significantly worsens the fit at the edges. The SED is minimally affected by the inner radius except for very large inner radii since this removes hot dust close to the star that is responsible for mid-infrared flux.
\end{itemize}

To quantify the quality of the fit for each set of parameters tested, we do a weighted linear least squares fit to the six data points in the SED as well as the six points in Fig.~\ref{rad_den} distributed along the sides of the 
brightness profile.  We avoid the central section of the brightness profile because 
our symmetric model cannot account for the asymmetric structure observed.  We choose six approximately spatially independent points in the brightness profile to give equal weight to the SED and image fitting.  
This test helps significantly in narrowing down the best values for most parameters, although it does not do a good job discriminating between values of $\alpha$.  

We have not done a complete parameter space study for the parameters listed above, so this least squares fit represents the best fit of the chosen parameter combinations.  However, we have attempted to choose values for the parameters 
such that the progression of tests 
becomes more and more constraining for each parameter.  Namely, starting with a coarse grid within the parameter space described in Sect.~\ref{params}, a group of parameters was tested. 
By learning the effect on the SED and brightness profile of varying an individual parameter, subsequent tests of parameter sets were designed to target those parameters that would compensate for the most deviant aspects of a preceding model.  This method narrowed down the parameter space and allowed finer sampling of the more realistic values.

\subsection{Comparison with T Tauri Stars}
Comparing IRAS 18151-1208 with CB 26 \citep{saut2009} and the Butterfly Star (IRAS 04302+2247, \citealp{wolf2003, wolf2008}), we notice that similar parameters describing the 
shape of the disk adequately reproduce the observed elongation in IRAS 18151-1208.  We did not choose these values right from the beginning, but only through quantifying the fits by means of calculating a weighted least squares 
value did we determine these parameters to be nearly identical.  

On absolute scales, the values of the disk scale height at 100 AU do not coincide (5 AU for IRAS 18151-1208 versus 10 AU for CB 26 and 15 AU for the Butterfly Star).  However, this discrepancy can be explained by the fact that in the T Tauri stars, 100 AU is proportionately much further out in the disk (ex.\ $\frac{1}{3}$R$_{\rm{out}}$ for the Butterfly Star) compared to the high mass protostellar object in which 100 AU corresponds to only one fiftieth of the outer radius ($\frac{1}{50}$R$_{\rm{out}}$).  As defined, these values cannot directly be compared.  As a better comparison, we calculate the height of the massive disk at $\frac{1}{3}$R$_{\rm{out}}$.  This turns out to be 200 AU, or 4\% of the outer radius.  In the CB 26 and Butterfly Star cases, the height of the disk one third of the way out is 3\% and 5\% of the outer radius respectively.  Within the uncertainties, these numbers are quite comparable and support our assumption that a geometrically thin disk is an appropriate model in the high mass regime.

The parameters describing the dust mass and the central star differ significantly for the two types of young stars as expected.  The mass of the modeled IRAS 18151-1208 system is on the order of 1000 times greater than in the T Tauri cases. Although the calculated mass may be an overestimate if there is a contribution to the millimeter flux from a dense envelope, we do not include an envelope in our models as that would increase the number of free parameters.  

Another large difference between the high mass protostellar object and T Tauri star systems is in the outer radius.  In the scenario that massive star formation proceeds as a scaled up version of low mass star formation, it is expected that disks around massive stars should be significantly larger than the disks of their low mass counterparts.  The observations in this case study are consistent with this picture, but definitive observational evidence of large disks around massive stars is lacking (See the review by \citealp{cesa2007}).  The elongated structure that we observe in IRAS 18151-1208 is more than ten times larger than the several hundred AU extent of disks around T Tauri stars and although we do not want to overinterpret the scaling-up nature of our results, it is interesting to note that the disk's radius and the mass of the disk (which would increase by the cube of the scaling factor) can be described by a scaling factor of $\sim$15 compared to the Butterfly Star. 

\subsection{Rotation and Infall} \label{rotation}

\begin{figure*}
\includegraphics[angle=-90,scale=0.7]{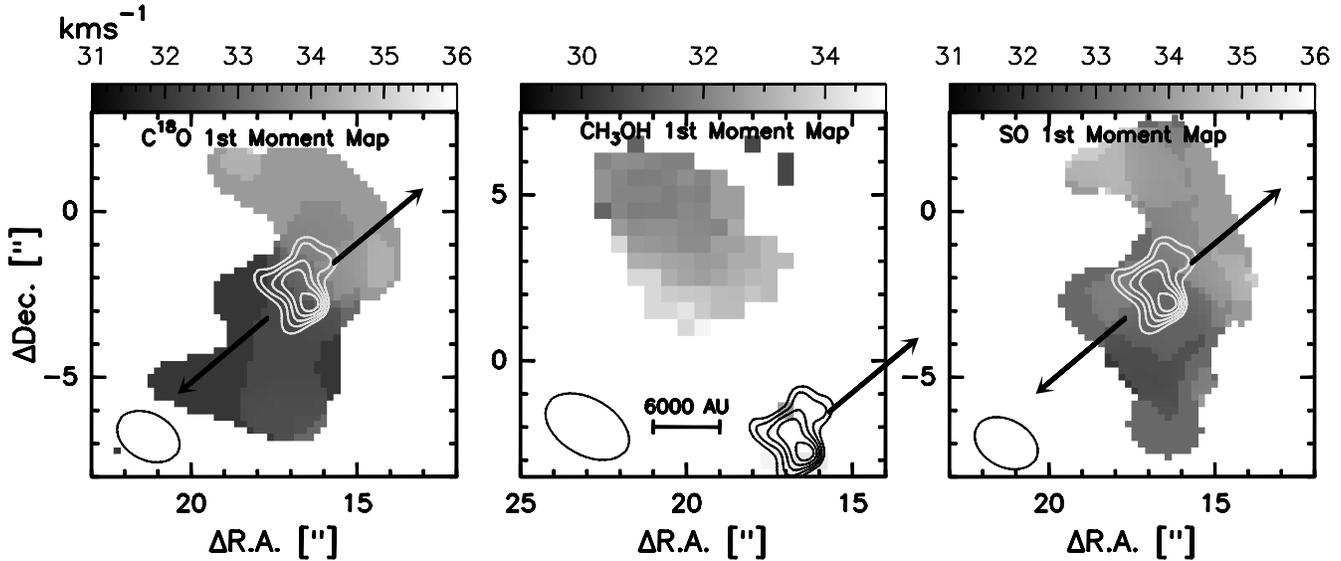}
\caption{C$^{18}$O, CH$_{3}$OH, and SO velocity moment maps clipped at the 8, 4, and 6$\sigma$ levels respectively of the corresponding line's intensity map.  For C$^{18}$O, $\sigma$=0.075 Jy, for CH$_{3}$OH, $\sigma$=0.054, and for SO, $\sigma$=0.062.  In each panel, the 
contours are those of the 1.3 mm dust continuum, the arrows indicate the orientation of the $^{12}$CO outflow, and the ellipse at bottom left represents the size of the SMA beam.  Although the size of each map is identical, note that the spatial position is offset in the middle panel compared to the other panels. \label{mom_maps}}
\end{figure*}

The scarcity of molecular lines in the observed spectrum has made it a difficult task to assess the region for indications of rotation or infall.  A velocity gradient in C$^{18}$O is mildly present (see Fig.~\ref{mom_maps}), but is neither aligned with the elongation in the expected disk orientation nor with the outflow direction.
This gradient has a velocity range of $\sim$4 kms$^{-1}$ and extends across a region approximately 25,000 AU long.  The SO emission follows a similar trend as C$^{18}$O, but the trend is even weaker.  
These molecules are likely affected by both the outflow and the disk component.

As for the other molecules detected in the SMA spectra (see Fig.~\ref{spectra}), $^{12}$CO and $^{13}$CO are well established outflow tracers.  The only other molecule detected in the region is methanol (CH$_3$OH).  However, 
there is no significant methanol detection at the position of the dust continuum peak, and the emission detected to the northeast does not exhibit any indication of rotation.  As there are no further lines in our 4 GHz bandwidth spectrum from the SMA, we are limited on this front.  Perhaps further observations at other wavelengths would yield a tracer of the kinematics in the region. 


\subsection{Predictions for Future Observatories} \label{future}

\begin{figure*}
\includegraphics[angle=-90,width=\textwidth]{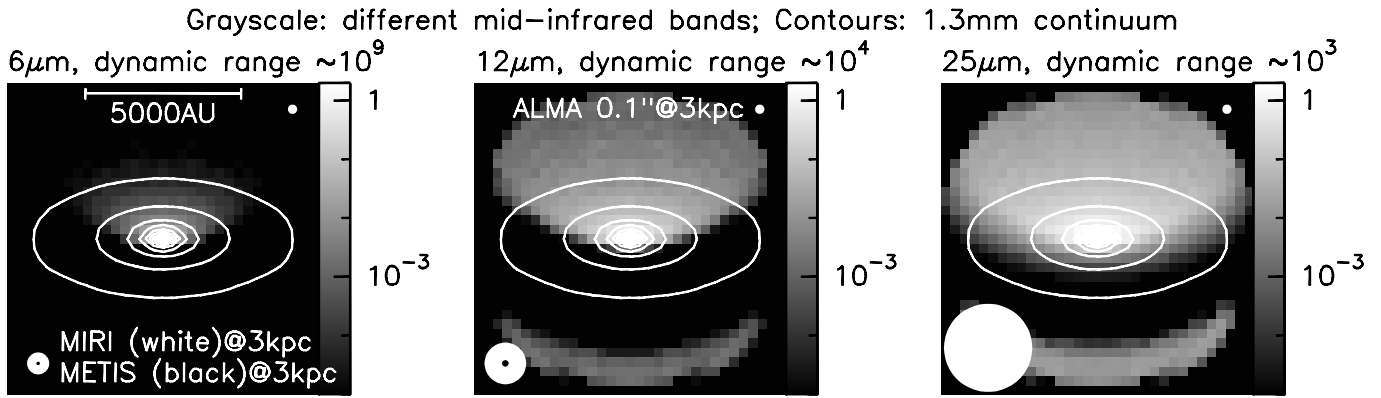}
\caption{Predictions at mid-infrared and millimeter wavelengths of what may be observable with future observatories. For the sake of example, resolution elements for several planned instruments (MIRI on JWST, METIS on the E-ELT, and ALMA) are shown in
each panel that correspond to wavelengths observable with 
the respective instrument.  The circles in the lower left corner correspond to the beam sizes of the MIRI (white circles) and METIS (black circles) instruments and the white circle in the upper right corners represents a 0.1$\arcsec$ ALMA beam. The flux scale is identical in all three panels. \label{prediction}}
\end{figure*}

The MC3D code is capable of computing images at a wide range of wavelengths which can be used to speculate what a disk might look like with future observatories.  In Fig.~\ref{prediction} we include predictions at 6, 12, and 25 $\mu$m as well as contour overlays of the 1.3 mm model.  In this figure, we also include the beam sizes of observatories on the horizon, namely the Atacama Large Millimeter Array (ALMA), and the MIRI and METIS instruments planned for the James Webb Space Telescope (JWST) and the European Extremely Large Telescope (E-ELT).  The flux is plotted on the same scale in each panel for the purpose of comparison.  It should be noted that despite increases in sensitivity over today's instruments, the dynamic range of these future observatories will still be insufficient for the dynamic range of the 6 $\mu$m image.
\section{Conclusions}

Observations of the high mass protostellar object IRAS 18151-1208 reveal an elongation in the 1.3 mm dust continuum that has a position angle 29$^{\circ}$ East of North.  These observations also confirm a well-defined orientation for a collimated massive outflow emanating from that dust continuum peak with a position angle 130$^{\circ}$ East of North.  This elongated structure is perpendicular to the outflow orientation, and has an approximate diameter of 10,000 AU.  While kinematic evidence of rotation is not available, the spatial signature of the dust continuum is indicative of a circumstellar disk.

Under the hypothesis that massive star formation is a scaled up version of low mass star formation, we adapted the 3D radiative transfer Monte Carlo code MC3D to reproduce the observations.  This was achieved by fitting the SED as well as the 
brightness profile along the major axis of the elongated 1.3 mm dust continuum image.  The best fit model is of a flared disk containing 220 M$_{\odot}$ of gas and dust with an inclination of 60$^{\circ}$. While this is not the only possible scenario to explain the observations, we conclude that this scaled up approach is viable.

Compared to disks around T Tauri stars for which MC3D has successfully been applied, our best fit model has similar parameters describing the shape of the disk (density distribution and amount of flaring) while larger values are necessary for the absolute parameters of the system such as mass, extent, and the central star.  The disk produced by the modeling is unstable aside from a small region close to the central star.  This is an expected outcome since the mass of the disk is dominant in the disk-star system.

A further outcome of the modeling is that we can produce images at multiple wavelengths which can be helpful for visualizing what future observatories such as ALMA, JWST, and the ELTs will be capable of observing.  Future work possible before these observatories become available includes looking for other tracers of the kinematics in the region.  Pending the discovery of a molecule that traces rotation in the region, this source will be unique in that both the spatial density and flaring structure as well as the kinematic properties of the region will be known.


\acknowledgments

C.F. and H.B. acknowledge support by the {\it Deutsche Forschungsgemeinschaft}, DFG project number BE 2578.  C.F. also acknowledges support from the {\it International Max-Planck Research School for Astronomy \& Cosmic Physics} at the University of Heidelberg.  J.S. acknowledges support by the DFG-Forschergruppe 759 ``The Formation of Planets: The Critical First Growth Phase.''  It is our pleasure to thank the anonymous referee for his/her comments leading to the improvement of our manuscript.


\bibliographystyle{aa}

\bibliography{library.bib}

\end{document}

%% file: acronyms.tex
\def\apjl{ApJ}%
\def\apjs{ApJS}%
\def\ao{Appl.~Opt.}%
\def\aap{A\&A}%